\documentclass[journal]{IEEEtran}
\usepackage[dvips]{graphicx}
\usepackage{amsmath, amsthm, amssymb, amsfonts, mathtools}
\usepackage{cases, multirow}
\usepackage{srcltx,color}
\usepackage{cite, url}
\usepackage{subfigure}

\long\def\comment#1{}

\def\figref#1{Fig.~\ref{#1}}
\def\be{\begin{equation} }
\def\ee{\end{equation} }

\title{\huge On Modeling Heterogeneous Wireless Networks Using Non-Poisson Point Processes}

\begin{document}
  

\author{
Young~Jin~Chun,	Mazen~Omar~Hasna, Ali~Ghrayeb, and Marco~Di~Renzo%
\IEEEcompsocitemizethanks{
  \IEEEcompsocthanksitem 
  Y. J. Chun is with the Wireless Communications Laboratory, ECIT Institute, Queens University Belfast, United Kingdom (Email: Y.Chun@qub.ac.uk).
  \IEEEcompsocthanksitem 
  M. O. Hasna is with Department of Electrical Engineering, Qatar University, Doha, Qatar (e-mail: hasna@qu.edu.qa).
  \IEEEcompsocthanksitem 
  A. Ghrayeb is with the Department of Electrical and Computer Engineering,
  Texas A\&M University at Qatar, Doha, Qatar (e-mail: ali.ghrayeb@qatar.tamu.edu).
  \IEEEcompsocthanksitem 
  M. D. Renzo is with the French National Center for Scientific Research (CNRS), 
  Paris, France, and also with the \'Ecole Sup\'erieure d'\'Electricit\'e, 
  University of Paris–Sud XI, Paris, France (e-mail: marco.direnzo@lss.supelec.fr).
  }
\thanks{This paper was made possible by Grant NPRP 4-1119-2-427
 from the Qatar National Research Fund (a member of Qatar Foundation). 
 The statements made herein are solely the responsibility of the authors.}
}

\maketitle


{\it \bf Abstract}-
{
Future wireless networks are required to support 1000 times higher data rate, than the current LTE standard. In order to meet the ever increasing demand for achieving reliable and ubiquitous coverage, it is inevitable that, future wireless networks will have to develop seamless interconnection between multiple technologies. A manifestation of this idea is the collaboration among different types of network tiers such as macro and small cells, leading to the so-called heterogeneous networks (HetNets). Researchers have used stochastic geometry to analyze such networks and understand their real potential. Unsurprisingly, it has been revealed that interference has a detrimental effect on performance, especially if not modeled properly. Interference can be correlated in space and/or time, which has been overlooked in the past. For instance, it is normally assumed that the nodes are located completely independent of each other and follow a homogeneous Poisson point process (PPP), which is not necessarily true in real networks since the node locations are spatially dependent. In addition, the interference correlation created by correlated stochastic processes has mostly been ignored. To this end, we take a different approach in modeling the interference where we use non-PPP, as well as we study the impact of spatial and temporal correlation on the performance of HetNets. To illustrate the impact of correlation on performance, we consider three case studies from real-life scenarios. Specifically, we use massive multiple-input multiple-output (MIMO) to understand the impact of spatial correlation; we use the random medium access protocol to examine the temporal correlation; and we use cooperative relay networks to illustrate the spatial-temporal correlation. We present several numerical examples through which we demonstrate the impact of various correlation types on the performance of HetNets, for both the PPP and non-PPP models. 
}

\section{Introduction}

During the last decade or so, the global smart-phone usage has grown explosively, 
driving the development of mobile communications faster than ever before. 
The first release of the Long Term Evolution (LTE) standard, Release 8, was finalized in 2008, the work on Release 12 was completed by 2014, 
and the fifth generation (5G) of mobile networks is expected to be deployed around 2020. 
Future wireless networks are expected to deliver 1000 times higher spectral efficiency than 4G to support more diverse set of devices reliably 
and ubiquitously. To reach higher spectral efficiencies, wireless network technologies need 
to collaborate and construct a seamless interconnection between multiple tiers, such as macro and small cells. 
Such interconnection between multiple tiers, which is normally referred to as heterogeneous networks (HetNets), 
has been shown to improve spectral efficiency and coverage \cite{ref01}.
However this may as well increase the in-band interference, leading to a drastic degradation in the overall networks performance. 

HetNets are inherently irregular and possess intrinsic randomness. For example, the deployment of small cells is generally based on user demand, 
which is irregular, implying that the deployment of base stations (BSs) is random as well. 
Fading, shadowing, path loss, transmit power and supported date rates are different for different tiers. 
The traffic distribution of various network architectures is different as well. 
As a consequence, analyzing and evaluating different communication technologies and access protocols over HetNets is very challenging \cite{ref01_ver2}. 
The most popular method of analyzing HetNets is to use the homogeneous Poisson point process (PPP) model, borrowed from stochastic geometry. 
This method models interference in wireless networks by treating the locations of the transmitters as points distributed according to a spatial Point Process. 
Such an approach captures the irregularity and variability of the node configurations found in real networks, 
provides theoretical insights, and allows high analytical flexibility with results that are as accurate as that of the traditional grid model. 

\subsection{Previous Work on PPP based Modeling: HIP Model}

PPP is suitable to model a network with infinitely many nodes that are randomly and independently distributed over a given service area.
Important properties of PPP include stationarity, \textit{i.e.}, its distribution is invariant under arbitrary translation, 
and the fact that the Slivnyak's theorem applies to it, which means that conditioning on a certain point does not 
change the distribution of the process. Due to these two properties, the PPP model is analytically tractable and flexible, 
which explains the pervasive popularity of the PPP model. 
Other appealing properties for HetNets are that the sum of PPPs is still a PPP, 
the probability generating functional (PGFL) is known in closed-form, 
and the distribution of the distance among the nodes is known, like the distance to the first or the $n$-th neighbor.
The probability density function of the aggregated interference, the Laplace transform of the interference,
and the outage probability were analyzed for a PPP network \cite{ref02}.

The homogeneous independent Poisson (HIP) model in \cite{ref10} assumes $K$ independent tiers of PPP distributed BSs for modeling HetNets,
and is by far the most well understood HetNets model in the literature due to its simplicity and tractability.
The PPP is useful for HetNets modeling since the sum of PPPs is still a PPP and the superposition of multiple PPPs 
can effectively model multi-tier HetNets. In \cite{ref10}, a tractable model for the signal-to-interference-plus-noise ratio (SINR) was proposed for a $K$-tier HetNet where the BSs are spatially distributed as PPP. 
The authors evaluated the coverage probability, the average load, and the average achievable rate 
for an unbiased cell association.

\subsection{Limitations of Previous Work}

While previous work have made significant advances from an analytical point of view, 
the HIP model fails to accurately model the real deployment of HetNets in two aspects. 
First, they assumed that the nodes are located independently of each other, \textit{i.e.}, no spatial dependence. 
However, in many actual wireless networks, node locations are spatially dependent, \textit{i.e.}, 
there exists repulsion (or attraction) between nodes, 
and the PPP assumption does not provide an accurate model for the interference in these conditions.

Second, regardless of using PPP or non-PPP, the adopted communication technologies and transmission protocols 
may introduce correlation in the aggregate interference, which were not properly reflected in the modeling process until very recently. 
Aggregated interference is spatially correlated since it originates from a single set of transmitters sharing common randomness.
It is also temporally correlated since a subset from the same set of nodes transmit in different time slots.
For example, spatial and temporal correlation occurs between different time slots if the receiver is equipped with a single antenna, 
between different receive antennas if the receiver is equipped with multiple antennas, 
or between different receivers if they are separated.

\subsection{Motivation}

These limitations motivate us to propose a more realistic model based on non-Poisson processes 
that accurately capture the non-idealities pertaining to the deployment of HetNets in real life applications.
In particular, we put forth a HetNet model that captures different types of correlations among different HetNet tiers. 
To substantiate the usefulness of this HetNet model, we investigate the impact of 
different types of correlations on the performance of HetNets. 
To be concrete, the following main case studies are considered for illustration:
\begin{enumerate}
 \item We use massive multiple-input multiple-output (MIMO) 
to illustrate the impact of spatial correlation on the outage probability performance. 
 \item We examine the temporal correlation on the local delay performance while using a random medium access protocol.
 \item We consider a cooperative relaying scenario where we examine 
the impact of spatial-temporal correlation on the packet delivery probability performance.
\end{enumerate}

We begin by proposing a flexible multi-tier architecture for HetNets, 
followed by an overview of the non-Poisson process and case studies to illustrate the impact of interference correlation on the non-Poisson process model.

\section{A Flexible Multi-Tier Architecture for HetNets: Beyond HIP Model}

\begin{figure}[!t]
    \centering
    \includegraphics[width=0.75\linewidth]{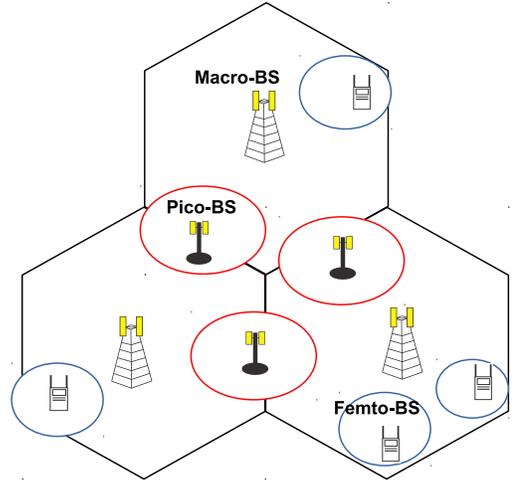}
    \caption{System model for Heterogeneous cellular networks.}
    \label{fig.system}
\end{figure}

Although this article targets HetNets with an arbitrary number of tiers, for illustration purposes, we consider a HetNet with three tiers, 
which was originally proposed in \cite{ref12}. The three tiers are described as follows and illustrated in \figref{fig.system}.
\begin{itemize}
 \item \textbf{Tier 1}: A macro-cell that is regularly distributed over the plane to provide wide area network coverage. 
 \item \textbf{Tier 2}: A pico-cell that is deployed near the cell boundaries of tier-1 to improve the network coverage or 
 in a highly populated building to give dedicated capacity to large individual groups. 
 \item \textbf{Tier 3}: A femto-cell that is an autonomous layer that adapts automatically to the underlying layers without any central planning.
It is usually installed by the end user to provide dedicated coverage and capacity to a small group. 
 \end{itemize}
This versatile model is general enough to capture various correlation types 
and flexible enough to model various scenarios for small cell deployment.
The specifications of the model depend on the particular type of scenarios that is used.

\begin{figure}[!t]
    \centering
    \includegraphics[width=0.8\linewidth]{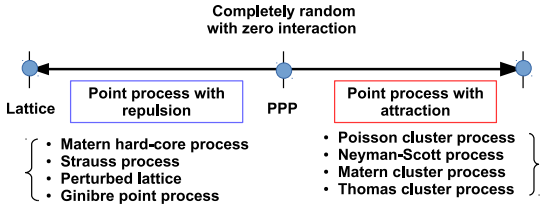}
    \caption{Stochastic dependencies amongst the point processes.}
    \label{fig.pp}
\end{figure}

\subsection{Point Process with Repulsion for Tier-1}

A point process with repulsion occurs when points are forbidden to be closer to each other than a certain minimum distance. 
In many actual cellular networks, two adjacent BSs can not be arbitrarily close to each other because of physical constraints and MAC scheduling. 
For MAC scheduling, the carrier sense multiple access (CSMA) protocol prevents two nearby nodes from transmitting at the same time. 
So spatial models that account for repulsion between transmitting nodes or other post-MAC correlations are required to model intelligent scheduling protocols. 

The following point processes have these repulsive features;
\begin{itemize}
 \item \textbf{Matern hard-core process:} 
  The Matern hard-core process can be generated from a homogeneous PPP by removing any points that are closer to each other than a predefined minimum distance $R$.
  Explicit formulas for its intensity, higher order product density and the pair correlation functions are derived in \cite{ref02_ver3}. However, only the approximate expressions of Laplace transform, characteristics function, or moment generating function are available for Matern hard-core process, limiting their applications. 
 \item \textbf{Determinantal point process (DPP):} 
DPP is a soft-core point process with adaptable repulsiveness; a hard-core process occurs by forcing an absolute minimum distance between any two nodes, 
whereas a soft-core process stems from assuming a weak repulsive force based on the inter-node distance.
Therefore, with small probability, some nodes can be closer than the minimum distance. 
DPP is a tractable point process where its Laplace transform and PGFL were derived in \cite{ref_ver3_01}. 
Based on these tools, the authors evaluated the distribution of the distance to the nearest neighbor node, mean interference 
and the signal-to-interference ratio (SIR) distribution and numerically proved that the DPP is a more accurate model for cellular network performance analysis than the PPP.
 \item \textbf{Ginibre point process (GPP):} $\beta$-GPP is a special case of DPP that has been recently introduced in \cite{ref18}.
  It can adaptively behave either as a PPP or a repulsive process, depending on a thinning parameter $\beta$;
  If $\beta = 1$, $\beta$-GPP is identical to the GPP, which is a repulsive point process. 
  If $\beta \rightarrow 0$, the $\beta$-GPP converges weakly to the PPP with intensity $1/\pi$. 
  Point distribution, Palm distribution, as well as higher order moment measures were evaluated in \cite{ref18}. 
  The mean and variance of the interference and the coverage probability of the typical user in an ad-hoc network
  are evaluated in \cite{ref18} using the $\beta$-GPP model.
 \end{itemize}

\subsection{Non-homogeneous PPP for Tier-2}

Non-homogeneous PPP allows the intensity of the point process to be location dependent, with function $\lambda(x)$, 
instead of the constant intensity $\lambda$ that corresponds to the PPP case. 
This class of point processes can model spatial variations of the trend that easily occurs in real world applications. 
For this reason, this process is appropriate for modeling the pico-BSs installed around the macro-cell boundaries to improve the network coverage.
The main difference between a non-homogeneous PPP and a repulsive (or cluster) process is that
non-homogeneous PPP occurs by the spatial variation of the intensity function, 
whereas the repulsive (or cluster) process occurs by the stochastic dependence amongst the points of the process. 
These two classes of processes have a fundamental ambiguity that makes it difficult to disentangle with each other. 
For example, non-homogeneous PPP can be used to model either repulsive or cluster process depending on the specific intensity measures. 

The following point processes have these non-homogeneous characteristics;
\begin{itemize}
  \item \textbf{Cox Process:} If the intensity measure of a Poisson process is random, 
the resulting process is conditioned on the realization of the intensity measure 
and it is  referred to as a spatial Cox process or doubly stochastic Poisson process. 
A Cox process is a general process that extends homogeneous PPP and is often used to model patterns 
exhibiting more clustering than PPP. Depending on the intensity measure, a Cox process can be specified as 
mixed PPP, random thinning process, Poisson hole process, and so on \cite{ref02}.
The functional form of Laplace transform and probability generating functional for a Cox process are available in the literature, however, 
the specific expression depends on the particular type of intensity measure that is used. 
  \item \textbf{Dependent Thinning Process:} A thinning process is obtained by applying a specific rule to delete or maintain each point of the basic point process.
If the thinning applies independently to the nodes of PPP without any interaction between the points, the resulting node locations are still
PPP and the resulting process is called an independent thinning process.
However, if the thinning depends on the inter-node distance or the number of adjacent nodes, then the obtained process is a dependent thinning process.
Explicit formulas for its intensity, second order product density and the pair correlation functions were derived in \cite{ref02}.
\end{itemize}

\subsection{Point Process with Attraction for Tier-3}

A point process with attraction, which is also known as a cluster process, occurs due to environmental conditions 
(office spaces, gathering spots, or highly populated cities) or intentionally induced by the MAC protocol. 
A cluster process is generated by a parent process that forms the center of a cluster 
and the daughter points that are spatially distributed around the cluster center. 
The cluster process is then the union of all the daughter points. 

Depending on the point processes of the parent and daughter points,
a cluster process can be classified as follows;
 if the parent points are distributed by PPP and the daughter points follow a general point process,
 then the cluster process is known as \textit{Poisson cluster process}.
 If the parent points follow a general point process, and the daughter points are distributed by PPP,
 then it is known as \textit{Cox cluster process}.
 If both the parent and daughter points are PPP distributed, 
 the it is know as \textit{Neyman-Scott process}.

  Two important special cases of the Neyman-Scott process are Matern and Thomas cluster processes; 
 \begin{itemize}
 \item \textbf{Matern cluster process:}
 If each daughter point is uniformly distributed in a circle of radius $R$ around the origin, the cluster process is categorized as Matern cluster process.
 \item \textbf{Thomas cluster process:} 
 If each daughter point is scattered around the origin based on a normal distribution with variance $\sigma^2$, the cluster process is known as Thomas cluster process.
  \end{itemize}
Although the statistics for Neyman-Scott processes are quite complicated, the analytical measures are well studied in the literature.
Explicit formula for its intensity, higher order product density and the pair correlation functions were derived in \cite{ref02_ver3}. 
Closed form expressions of Laplace transform, probability generating functional, and moment generating functions were evaluated for the Neyman-Scott process in \cite{ref19}. The outage probability and average achievable rate of HetNets were evaluated for Poisson cluster process in \cite{Chun2015}.

\section{Correlation in the point process: Case Studies}

The key questions that arise while using non-Poisson processes are the interference correlation and the joint distribution of the interference. 
For non-Poisson processes, the number of points in disjoint regions may be dependent on each other. 
This means that conditioning on a point being at a certain location changes the statistics of the point process. 
As a result, correlations and higher-order statistics of the interference become important while analyzing non-Poisson processes. 
While it has been long recognized that correlated fading degrades the performance gain in multi-antenna communications,
interference correlation has been ignored until very recently \cite{ref22}. Interference correlations generally originate from
either the spatial distribution of transmitters or the protocols/transmission technologies being used (like relaying and multiple-antennas), 
since they determine the locations and the active pattern of the interferer, which in turn decides the structure of the interference. 
Also, it is important to note that the interference correlation occurs for both PPP/non-PPP, 
but it is more emphasized and severely affects the system performance for non-PPP.
The interference correlation can be classified into three categories based on the receiver configurations; 
\textbf{1) correlation between different receive antennas}, 
\textbf{2) correlation between different time slots at the same receiver},
and \textbf{3) correlation between different receivers}. 
In the following case studies, we introduce each class of correlation and describe how this impacts the network performance. 

\subsection{Case 1. Spatial Correlation: Massive MIMO}

Massive MIMO is a transmission technique that uses an extremely large number of antennas to support multiple users. 
Massive MIMO can aggressively use spatial multiplexing to increase the capacity 10 times or more, 
while simultaneously improving the radiated energy-efficiency by orders of magnitude. 
In the massive MIMO, interference correlation between different receive antennas occurs, \textit{i.e.}, spatial correlation,
because the interference is originated from the same source of transmitters.

The spatial correlation has an impact on the diversity order. Since the interference powers at each receive antenna are correlated, 
the SINRs at the antennas are not independent, and the diversity is smaller than generally assumed. 
The authors in \cite{ref28} proved that the success probability of spatially correlated SIMO link is proportional to 
$\log P_M \propto M^{2/\alpha}$ as opposed to $\log P_M \propto M$ for independent interference, 
where $\alpha$ is the path-loss exponent and $M$ is the number of receive antennas. 
This suggests that the effect of spatial correlation becomes more evident as the path-loss exponent increases
and this effect should be accounted for in analyzing the performance of massive MIMO.

One efficient method to suppress the spatial correlation is to use interference aware MRC combining at the massive MIMO receiver. 
Conventional MRC combining weights are proportional only to the channel conjugate 
and do not depend on the interference power experienced at each antenna. 
If the number of receive antennas is large, one of them might be suffering from an impulsive interference event,
which can corrupt the entire detection process and worsen the outage performance. 
Instead, the authors in \cite{ref29} proposed a MRC receiver that chooses a weight vector directly proportional to the channel conjugate 
and inversely proportional to the interferer density corresponding to the interference field seen by each antenna. 
Since the interferer density is proportional to the average interference power, 
this method can adapt to the interference and effectively suppress the spatial correlation.

\subsection{Case 2. Temporal Correlation: MAC Protocols}

The aggregate interference in HetNets is a stochastic process in which randomness originates from the spatial node distribution, MAC protocol behavior, and random channel fading. At the receiver of each tier, the interference correlation between different time slots occurs, \textit{i.e.}, temporal correlation. The interferences at two different time slots are correlated because they come from correlated sets of transmitters and the resulting traffic may also be correlated. For example, denote $A_k$ as the event that a link outage between the transmitter and receiver occurs during time instant $k$. The authors in \cite{ref22} proved that $P(A_k|A_l) > P(A_l)$, \textit{i.e.}, if a link outage occurred during time slot $l$, there is a significant probability that the link outage will also occur at time $k$, where $k > l$. Therefore, blindly retransmitting, while ignoring the temporal correlation will simply reduce the transmission rate and increase the local delay within the communication link. 

One effective method to reduce the temporal correlation is to intentionally induce man-made randomness by using random medium access; \textit{i.e.},
increasing randomness in the MAC domain, specifically frequency-hopping multiple access (FHMA) and ALOHA. In FHMA, the entire frequency band is divided into $N$ sub-bands, and each transmitter randomly chooses a sub-band in each time slot. 
Similarly, in ALOHA, each node transmits with a certain probability $p$ using the entire frequency band. 
These mechanisms increase the uncertainty in the active pattern of interfering nodes and reduce the intensity of the interfering transmitters, 
thereby reducing the effect of interference correlation. 
In \cite{ref26}, the authors analyzed the time to transmit a packet from a node to its intended receiver, \textit{i.e.}, local delay, in FHMA and ALOHA based on PPP. If no MAC dynamics is employed, the local delay has a heavy tail distribution which results in an infinite mean local delay; meanwhile, employing FHMA and ALOHA greatly decreases the mean local delay. By analyzing the local delay, the authors determined the optimal number of sub-bands in FHMA and the optimal transmit probability in ALOHA that minimizes the mean local delay.

\subsection{Case 3. Spatio-Temporal Correlation: Relaying}

Cooperative relaying in HetNets is an effective technique for improving the reliability and throughput of the total network and 
has attracted much attention in both academia and industry.  
For the cooperative schemes in HetNets, different user equipments (UEs) and eNodeBs (eNBs) are allowed 
to share resources and channel information to implement collaboration.

In cooperative relaying, the interference correlation occurs between different receivers that are closely located to each other. 
In \cite{ref30}, the authors assumed the PPP model for the interfering nodes and proved that the temporal and spatial correlation of 
interference significantly degrades the performance of the cooperative relay. 
They considered the effect of interference correlation for various link qualities, different relay locations, several detection strategies at the destination. 
In particular, when interference has high temporal and spatial correlation, the outage probability increases 
more than in the uncorrelated case, especially when the link qualities are good and relays are located close to the destination.
For the correlated case, the interference power at relay $k$ is similar to that received at other relays and it reduces the relay selection gain. 
This effect is particularly highlighted when the source-to-relay link is weak, \textit{i.e.}, the relays are closer to the destination. 
The opposite behavior is observed when the link qualities are bad. 
Therefore, the temporal and spatial characteristics of the interference significantly affect the system performance of cooperative relaying in HetNets, as well as the link qualities, relay locations, and detection algorithm.

\subsection{Numerical Results}

\begin{figure}[!t]
    \centering
    \includegraphics[width=0.85\linewidth, height=0.7\linewidth]{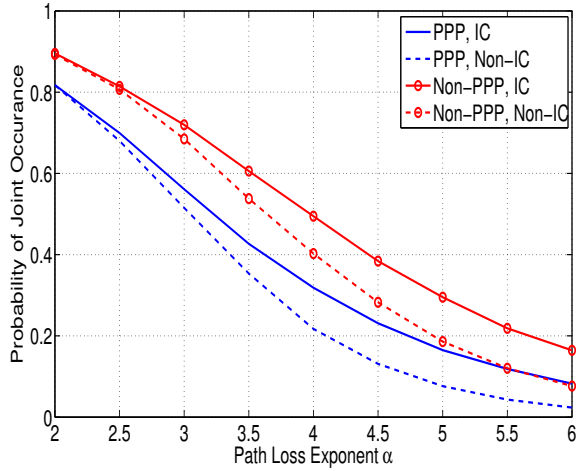}
    \caption{Probability of joint occurrence versus the path loss exponent, illustrating the impact of the spatial correlation.}
    \label{fig.spatial}
\end{figure}

In this section, we present numerical examples through which we compare the performance of PPP and non-PPP. 
We consider a three-tier HetNet where each tier's BSs are spatially distributed over a two-dimensional square.
We assume that the mobile of interest is associated to the macro-cell BS at the origin 
and the Matern hard-core process is proposed as a non-PPP model, \textit{i.e.}, a repulsive process.
\figref{fig.spatial} shows the joint occurrence probability of a $1 \times 2$ SIMO uplink versus the path loss exponent $\alpha$,
where the event of joint occurrence is defined as the complementary event of outage, \textit{i.e.}, $P(\text{joint occurrence}) = 1 - P(\text{outage})$ \cite{ref28}.
The blue and red lines are the joint occurrence probability of PPP and non-PPP, respectively, and 
the solid and dotted lines are the joint occurrence probability of interference correlated (IC) and interference uncorrelated (non-IC), respectively.
Since the Matern hard-core process excludes the points that are closer than a certain minimum distance, 
the probability that an interfering BS being close to a mobile user is smaller for non-PPP than that for PPP,
achieving a higher joint occurrence probability and a lower outage probability.
We also observe an evident performance gap between the correlated and uncorrelated interference cases for both PPP and non-PPP, 
and this gap increases for a larger path loss exponent $\alpha$.

\begin{figure}[!t]
    \centering
    \includegraphics[width=0.85\linewidth, height=0.7\linewidth]{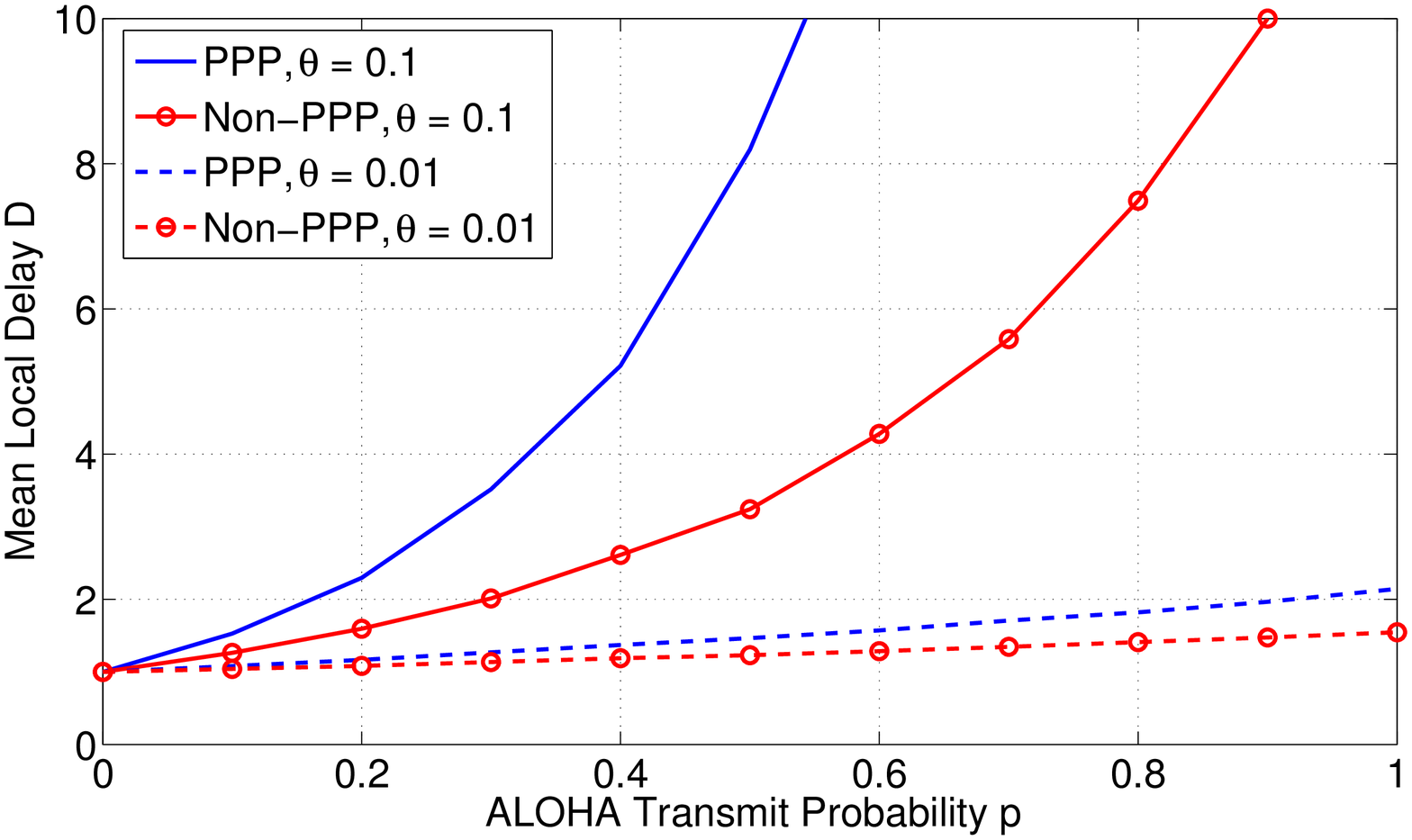}
    \caption{Mean local delay versus the ALOHA parameter $p$, illustrating the impact of the temporal correlation.}
    \label{fig.temporal}
\end{figure}

\figref{fig.temporal} compares the mean local delay versus ALOHA transmit probability $p$ for PPP and non-PPP. 
The mean local delay is defined in \cite{ref26} as the average number of required time slots for successful transmission.
Since the required time slots for successful transmission can be modeled as a geometric distribution with success probability 
$P(\text{success}) = 1 - P(\text{outage})$ and conditioned on the location of mobile, 
the mean local delay can be derived as the inverse of success probability averaged over the mobile location. 
In ALOHA, decreasing the transmit probability reduces the aggregate interference at the origin, thereby reducing the mean local delay.
Since the Matern hard-core process guarantees a minimum separation between interfering nodes, 
the success probability of non-PPP is larger than that of PPP, leading to a lower mean local delay. 

\begin{figure}[!t]
    \centering
    \includegraphics[width=0.85\linewidth, height=0.7\linewidth]{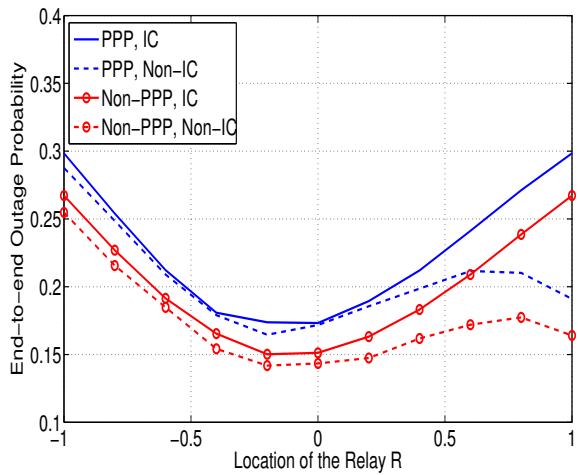}
    \caption{End-to-end outage probability versus the relay location, illustrating the impact of the spatial-temporal correlation.}
    \label{fig.spatial-temporal}
\end{figure}

\figref{fig.spatial-temporal} compares the end-to-end outage of a relay network versus the location of relay node. 
We use selection combining for detection at the destination and 
assume that the source node is located at $[-1, 0]$, the destination node is deployed at $[1, 0]$, and 
the relay node is positioned at $[R, 0]$, where $-1 \leq R \leq 1$.
We observe that non-PPP achieves lower outage probability than PPP, due to the guaranteed minimum separation between interfering nodes.
We also note that the interference correlation increases the outage probability with respect to the uncorrelated case, 
especially when the relay is located closer to the destination. If the relay moves closer to the destination, 
the interference values at the destination and the relay are likely to be very similar
in the correlated case. This reduces the selection combining gain and increases the outage probability, 
similar to what was observed in \cite{ref30}.

\section{Discussion and open issues}

So far, most of the related works on stochastic geometry assumed an idealistic PPP model without interference correlation 
and focused on simple transmission protocols, such as point-to-point transmission without re-transmission. 
However, as wireless networks evolve toward an interconnected multi-tier structure, 
the location of nodes will be spatially dependent on each other and the generated interferences will be correlated in space and time, 
so that the traditional approach can not provide an accurate model for real networks. 
These new trends of future wireless network introduce a number of open problems that must be addressed.

First, a comprehensive study on the properties of non-PPP must be carried out. 
Extensive research on PPP and Poisson cluster processes has already been done and tractable results 
on the interference distribution, outage probability, and transmission capacity of each model were established. 
In contrast, the Ginibre point process was proposed recently as a repulsive process 
and only the mean and variance of the interference, and the coverage probability were evaluated. 
Higher order moment measures of non-PPP are open problems that need to be derived or simplified for analytical tractability. 

Second, another important open problem is the impact of interference correlation on non-PPP networks. 
Most research on interference correlation assumed PPP and analyzed the loss of diversity or local delay due to correlated interference. 
Since the correlations and higher-order statistics of the interference significantly affect the communication performance of the non-PPP model, the local delay and diversity polynomial, which were analyzed based on homogeneous PPP, 
should be generalized to the non-PPP model.

Finally, efficient mechanisms to suppress the interference correlation over random wireless networks need to be developed.
The randomness of interference processes originates from the spatial node distribution, MAC protocol behavior, and random channel fading. 
As described in this manuscript, one effective method to suppress the temporal correlation is to induce man-made randomness by using random medium access. 
The artificially induced uncertainty in the active pattern of interfering nodes reduces the effect of interference correlation. 
However, the artificial uncertainty can also be implanted in either the spatial distribution or channel fading.  

These open issues provide future research directions in stochastic geometry that may help to successfully implement future wireless networks
while achieving the desired objectives.

\section{Conclusion}

In this article, we attempted to provide a high-level overview of the non-Poisson point processes
and to give a concise summary of the recent research trends on interference correlation. 
In particular, we proposed a realistic model for heterogeneous cellular networks based on non-Poisson processes 
that accurately captures the non-idealities pertaining to the deployment of HetNets in real life. 
We studied the impact of interference correlation on the performance of HetNets through case studies; 
we examined spatial correlation in massive MIMO link, temporal correlation in random medium access protocol, 
and spatial and temporal correlation in cooperative relaying.
We also performed a high-level overview of the existing literature on these topics
and demonstrated the performance degradation due to the presence of correlated interference in both PPP and non-PPP models. 
We have also identified a number of open research problems related to stochastic geometry with applications to HetNets.


 \bibliographystyle{IEEEtran}
 \bibliography{correl2}


\end{document}